\documentstyle[twocolumn]{article}
\newcommand{\beq}{\begin{equation}}
\newcommand{\eeq}{\end{equation}}
\newcommand{\bea}{\begin{eqnarray}}
\newcommand{\eea}{\end{eqnarray}}

\def\nuc#1#2{\relax\ifmmode{}^{#1}{\protect\text{#2}}\else${}^{#1}$#2\fi}
\title{\large
\vspace{-0.2cm}
EXPERIMENTAL AND COMPUTER SIMULATION STUDY OF RADIONUCLIDE PRODUCTION
IN HEAVY MATERIALS IRRADIATED BY INTERMEDIATE ENERGY PROTONS}
\vspace{0.5cm}
\author{Yu. E. Titarenko, O. V. Shvedov, 
V. F. Batyaev, E. I. Karpikhin, V. M. Zhivun,\\
R. D. Mulambetov, A. N. Sosnin\\
{\it Institute for Theoretical and Experimental Physics,
B.Cheremushkinskaya 25,}\\ {\it 117259 Moscow, Russia,
e-mail: Yury.Titarenko@vitep5.itep.ru} \vspace{0.1cm} \\
S. G. Mashnik, R. E. Prael \\
{\it Los Alamos National Laboratory, Los Alamos, NM 87545, USA}\vspace{0.1cm}\\
T. A. Gabriel\\
{\it Oak Ridge National Laboratory, Oak Ridge, TN 37831, USA} \vspace{0.1cm}\\
M. Blann\\
{\it 7210E Calabria ct, San Diego, CA 92122, USA}}\date{}
\begin{document}
\maketitle

\vspace{-0.0cm}
\noindent ABSTRACT
\vspace{0.1cm}

The results of measurements and computer simulations 
are presented for the yields of residual product nuclei in thin targets:
\nuc{nat}{U} irradiated by 0.1, 0.8, 1.2, and 1.6 GeV and \nuc{99}{Tc} 
irradiated
by       0.2, 0.8, 1.0, 1.4, and 1.6 GeV protons.
The yields  were
measured by direct high-precision $\gamma$-spectrometry.
The irradiations were made using the beams extracted from the ITEP U-10 
synchroton. The majority of these yields are measured for the first time at such 
energies.

We think that these new data on high energy fission yields measured together
with spallation and
fragmentation products are of importance both for development of new reliable 
models
of intermediate energy nuclear reactions and to benchmark available codes used 
today in applications.

About 820 cross sections are presented and used in comparison between measured 
yields and simulations by the  LAHET, INUCL, CEM95, HETC,
CASCADE, YIELDX, and ALICE codes.

\vspace{0.3cm}
\noindent I. INTRODUCTION
\vspace{0.1cm}

In designing ADT-based systems the neutron production of the 
Pb-Bi or Hg targets in the systems are of great importance and involve 
not only spallation,
% reactions 
but also fission reactions. The 
absence of any reliable theory for fission induced by 
100-2000 MeV protons necessitates experimenting with a broad range of 
nuclei, in our case with \nuc{nat}{U}. The data thus obtained may 
be used to construct semi-empirical systematics, to develop models
of nuclear reactions, and 
to verify available 
simulation codes. At the same time, the data are interesting by 
themselves because some present-day projects (see, for instance, 
Ref. $^1$) include actinides as part of the target materials.

The \nuc{99}{Tc} isotope 
(T$_{1/2} = 2.1\cdot10^5$ years) is a fission product whose yield 
($\sim$6\%) belongs to the extremely-high yield range and, therefore, 
has to be transmuted into stable or short-lived isotopes. One of the 
feasible techniques for transmuting \nuc{99}{Tc} is to add this isotope 
to the target materials of subcritical facilities whose neutron sources 
are neutron cascades generated in 0.8-1.6 GeV proton beam interactions 
with the targets.

Since (p,x) reaction data have yet to be obtained for \nuc{99}{Tc} 
(see Ref. $^2$), the feasibility of the given \nuc{99}{Tc} transmutation 
technique must be determined by experimenting with, and simulating, 
the \nuc{99}{Tc} interactions with protons of energies from 50--100 MeV 
to $\sim$3 GeV. Also, some of the \nuc{99}{Tc} reaction product yields 
are of interest in medical applications.

The present work was aimed at:
\begin{itemize}
\vspace{-0.1cm}
\item finding the independent and cumulative yields of residual product 
 nuclei in thin \nuc{nat}{U} targets irradiated with 
0.1, 0.8, 1.2, and 1.6 GeV protons, and in 
\nuc{99}{Tc} targets irradiated 
with  0.2, 0.8, 1.0, 1.4, and 1.6 GeV protons;
\vspace{-0.08cm}
\item 
comparing experimental data 
with results of simulations by the most extensively used codes that 
simulate the neutron-physics parameters of the subcritical systems 
mentioned above.
\end{itemize}

\vspace{-0.1cm}
Some of our 
results can be found in~Refs.~$^{3,4}$.

\vspace{0.4cm}
\noindent II. EXPERIMENTAL TECHNIQUES AND MEASURED YIELDS
\vspace{0.1cm}

The complete cycle of experiments have been made at  \linebreak

\begin{table*}%[b]
\caption{Irradiation parameters, sample dimensions and 
\nuc{27}{Al}(p,x)\nuc{22}{Na} reaction cross sections.} \label{tab1}

\vspace*{3mm}
%\begin{center}\begin{tabular}{|c|c|c|c|c|}\hline
\begin{center}\begin{tabular}{ccccc}\hline
Planned proton& Actual proton &\nuc{27}{Al}(p,x)\nuc{22}{Na} &
Sample/monitor & Total proton flux \\ 
energy &energy &reaction cross &thickness &across a sample for\\ 
(MeV) &(MeV)&section (mb)&(mg/cm$^2$)& irradiation time (p/cm$^2$)\\ \hline
& & & & \\
\multicolumn{5}{c}{ \nuc{99}{Tc} irradiations} \\ \hline
 200& 196.3 $\pm$ 1.1& 16.2 $\pm$ 1.2& 56/196& 1.62$\cdot10^{13}$\\ %\hline
 800& 795 $\pm$ 1& 15.4 $\pm$ 1.0& 67/142& 4.04$\cdot10^{13}$\\ %\hline
 1000& 989 $\pm$ 1& 14.9 $\pm$ 1.0& 55/195& 1.95$\cdot10^{13}$\\ %\hline
 1400& 1379 $\pm$ 1& 13.8 $\pm$ 1.0& 56/58& 5.92$\cdot10^{13}$\\ %\hline
 1600& 1583 $\pm$ 3& 13.3 $\pm$ 1.0& 51/58& 2.08$\cdot10^{13}$\\ \hline
% 1200& 1186 $\pm$ 1& 11.3 $\pm$ 0.8& 55/215& 1.23$\cdot10^{13}$\\ \hline
% 150& 146.5 $\pm$ 0.8& 10.9 $\pm$ 1.1& 60/197& 3.78$\cdot10^{12}$\\ \hline
% 100& 96.5 $\pm$ 0.6& 11.0 $\pm$ 1.2& 65/188& 6.68$\cdot10^{12}$\\ \hline
% 2600& 2574 $\pm$ 7& 11.0 $\pm$ 0.8& 56/190& 3.13$\cdot10^{12}$\\ \hline
& & & & \\
\multicolumn{5}{c}{ \nuc{nat}{U} irradiations} \\ \hline
 100& 96.5 $\pm$ 0.6& 19.5 $\pm$ 2.0& 55/183.5& 7.35$\cdot10^{12}$\\ %\hline
800& 795 $\pm$ 1& 15.4 $\pm$ 1.0& 55/185.8& 5.47$\cdot10^{13}$\\ %\hline
1200& 1186 $\pm$ 1& 14.3 $\pm$ 1.0& 55/215& 6.11$\cdot10^{13}$\\ %\hline
1600& 1583 $\pm$ 3& 13.3 $\pm$ 1.0& 51/58& 4.37$\cdot10^{13}$\\ \hline
 \end{tabular} \\[0.5ex]
\end{center} 
\end{table*}
\noindent{
ITEP for two years 
using the U-10 synchrotron with two 
proton beams extracted independently, 
namely, a high-energy beam with 800--2600 MeV extracted protons and a 
low-energy beam with 100--200 MeV extracted protons.}
 
The background $\gamma$-lines, which are present in measured 
$\gamma$-spectra due to natural radioactive background of the workroom,
to $^{nat}$U fission products, and to possible radioactive impurities in 
samples, were allowed for by 
analyzing $\gamma$-spectra of the intact samples measured under the same 
measurement conditions as those for the irradiated samples.
 
The $\oslash$10.5 mm metallic Tc foils were irradiated. The 
\nuc{27}{Al}(p,x)\nuc{22}{Na} monitor reaction was used in the 
present work.
Table \ref{tab1} presents the irradiation parameters, sample dimensions, 
and the \nuc{27}{Al}(p,x)\nuc{22}{Na} reaction cross sections.

The techniques and results
of measuring $\gamma$-ray spectra, processing $\gamma$-ray
spectra, determining external proton beam energies,
measuring geometry parameters of external proton beam shapes,
determining neutron background,
as well as the method for determination of radioactive nuclide yields
are described in detail in Ref.$~^{5}$. %\cite{cocu}.

Tables \ref{tc} and \ref{uran} present the products measured in 
\nuc{99}{Tc} and \nuc{nat}{U}.

\vspace{0.5cm}
\noindent III. COMPUTER SIMULATION OF MEASURED PRODUCTS
\vspace{0.2cm}

The products of the studied reactions were simulated 
in the present work by eight different codes,
namely:
%\begin{quote}
%\begin{itemize}
\begin{enumerate}
\vspace*{-0.2cm}
\item the CEM95 Cascade-Exciton Model code$^{6}$, %\cite{cem},
\vspace*{-0.3cm}
\item the CASCADE cascade -- evaporation -- fission transport
code$^{7}$, %\cite{cascade},
\vspace*{-0.3cm}
\item the INUCL cascade -- preequilibrium -- evaporation -- fission
code$^{8}$, %\cite{inucl},
\vspace*{-0.3cm}
\item the HETC cascade -- evaporation transport code$^{9}$, % \cite{hetc},
\vspace*{-0.6cm}
\item the LAHET cascade -- evaporation -- fission transport code$^{10}$,
%\cite{lahet},
%\vspace*{-0.3cm}
%\item the GNASH code based on the Hauser-Feshbach and preequilibrium 
%approaches$^{11}$, %\cite{gnash},
\vspace*{-0.3cm}
\item the ALICE code with HMS precompound approach$^{11}$, 
%\cite{alice96},
%\vspace*{-0.3cm}
%\item the Quantum Molecular Dynamics (QMD) code$^{19}$, %\cite{qmd},
%\vspace*{-0.6cm}
%\item the NUCLEUS cascade -- evaporation -- fission code$^{20}$, %\cite{tak1},
\vspace*{-0.3cm}
\item the semi-phenomenological YIELDX code$^{12}$, %\cite{yield},
\vspace*{-0.3cm}
\item the semi-phenomenological formulae of Foshina et al.$^{13}$ 
%\cite{foshin}.
%\end{itemize}
\end{enumerate}
%\end{quote}

A detailed description of the models used by  the CEM95,
CASCADE, INUCL, HETC,
LAHET, ALICE, and YIELDX
%, and GNASH 
codes
may be found in our previous work$^5$, %\cite{bi_nim}, 
in Refs.$^{6-12}$
%~\cite{cem}--\cite{alice96} 
and references therein.
%The  QMD, NUCLEUS, and
%YIELDX codes and formulae by Foshina et al. are briefly described in$^{10}$ 
%%\cite{cocu}
%and may be found in Refs.$^{19-22}$  %~\cite{qmd}-\cite{foshin}.

%In the present work,
The comparisons of measured data with the calculations
are made using two parameters.
The first parameter is numbers of ``coincidences". A coincidence is defined
to be a comparison event with the ratio
of simulated to experimental yields not 
exceeding a factor of 2
($N_{C_{2.0}}: ~~ 0.5 < \sigma_{cal,i} / \sigma_{exp,i} < 2.0$),
and a factor of 1.3
($N_{C_{1.3}}: ~~ 0.769 < \sigma_{cal,i} / \sigma_{exp,i} < 1.3$).
We present here the ratios of coincidence numbers
($N_{C_{1.3}}, N_{C_{2.0}}$)
 to the number of all comparisons ($N_S$).
The number of the coincidences within a factor of 1.3 ($N_{C_{1.3}}$)
is considered here to be the number of simulations within 
an ultimate accuracy of 30\% needed in applications$^2$. %\cite{Konin}.

Another parameter for comparison of simulated and experimental data
was proposed by R. Michel$^{14}$ %\cite{Mich3} 
and used afterwards in our works.$^{5,15}$ % \cite{bi_nim}.
The parameter is a mean squared deviation factor
\begin{equation}                                                    % (36)
\label{36}
\langle H \rangle = 10^
{\sqrt{\displaystyle \langle\left(lg\left( \frac{\sigma_{cal,i}}{\sigma_{exp,i}}
\right)\right)^2\rangle}} \mbox{ ,}
\end{equation}

\begin{table*}%[t]
\caption{Experimental yields measured in \nuc{99}{Tc}.}
\label{tc} \begin{center}

 \begin{tabular}{lccccccc}
 \hline Product& Half life & Mode& \multicolumn{4}{c}{Proton enegy (GeV)}\\ 
\cline{4-8} & & & 0.2 & 0.8 & 1.0 & 1.4 & 1.6\\ \hline 
\nuc{97}{Ru}& 2.9d  & ind & 13.1$\pm$ 1.4 & 2.57 $\pm$0.21& 3.34 $\pm$0.42& 2.22 $\pm$0.19& 1.57$\pm$0.17\\
\nuc{95}{Ru}&  1.643h & ind & 6.08$\pm$0.61 & 4.91 $\pm$1.10& 8.20 $\pm$1.16& 4.77 $\pm$0.87& 4.64$\pm$1.28\\
\nuc{94}{Ru}&  51.8m  & ind & 2.64$\pm$0.90 & 0.52 $\pm$0.21& 0.66 $\pm$0.29& 0.26 $\pm$0.18& 0.61$\pm$0.29\\
\nuc{99}{Tc}&  6.01h  & cum & 5.87$\pm$0.61 & 5.60 $\pm$0.47& 6.41 $\pm$0.53& 6.13 $\pm$0.55& 7.37$\pm$0.78\\
\nuc{96}{Tc}&  4.28d  & ind & 51.0$\pm$ 4.9 & 22.6 $\pm$1.7 & 21.1 $\pm$1.7 & 23.4 $\pm$1.9 & 19.5$\pm$1.8 \\
\nuc{95}{Tc}&  20.0h  & cum & 51.4$\pm$ 5.3 & 19.1 $\pm$1.4 & 17.2 $\pm$1.3 & 15.5 $\pm$1.3 &  \\
\nuc{95}{Tc}&  20.0h  & ind & 44.3$\pm$ 4.7 & 14.2 $\pm$1.5 & 9.02 $\pm$1.21& 10.7 $\pm$1.2 &  \\
\nuc{94}{Tc}&  52.0m  & cum & 11.5$\pm$ 1.2 & 3.25 $\pm$0.27& 2.57 $\pm$0.26& 2.68 $\pm$0.26& 2.22$\pm$0.26\\
\nuc{94}{Tc}&  52.0m  & ind & 7.70$\pm$0.87 & 2.74 $\pm$0.34& 1.91 $\pm$0.35& 2.42 $\pm$0.30& 1.61$\pm$0.31\\
\nuc{94}{Tc}& 293m  & cum & 28.0$\pm$ 2.7 & 9.67 $\pm$0.70& 8.57 $\pm$0.61& 7.67 $\pm$0.65& 7.21$\pm$0.69\\
\nuc{94}{Tc}& 293m  & ind & 35.7$\pm$ 2.9 & 12.4 $\pm$0.8 & 10.8 $\pm$0.7 & 10.1 $\pm$0.7 & 8.82$\pm$0.75\\
\nuc{93}{Tc}&  43.5m  & ind & 1.36$\pm$0.43 & 2.08 $\pm$0.86& 0.89 $\pm$0.49& 1.58 $\pm$0.50& 2.38$\pm$0.62\\
\nuc{93}{Tc}&  2.75h  & cum & 21.3$\pm$ 2.2 & 5.68 $\pm$0.61& 4.37 $\pm$0.35& 4.60 $\pm$0.40& 4.35$\pm$0.43\\
\nuc{93}{Tc}&  2.75h  & ind & 18.5$\pm$ 2.1 & 3.92 $\pm$1.08& 3.48 $\pm$0.60& 3.02 $\pm$0.61& 1.99$\pm$0.69\\
%\nuc{99}{Mo}&  65.94h & cum & & 13.6 $\pm$2.5 & & & 55.7$\pm$22.7\\
\nuc{93}{Mo}&  6.85h  & ind & 10.8$\pm$ 1.1 & 5.48 $\pm$0.38& 3.93 $\pm$0.64& 4.38 $\pm$0.39& 3.69$\pm$0.47\\
\nuc{90}{Mo}&  5.67h  & cum & 7.23$\pm$1.08 & 3.90 $\pm$0.30& 3.16 $\pm$0.53& 2.70 $\pm$0.27& 2.51$\pm$0.41\\
\nuc{96}{Nb}&  23.35h & ind & 1.18$\pm$0.18 & 2.03 $\pm$0.15& 2.70 $\pm$0.37& 2.39 $\pm$0.21& 2.33$\pm$0.29\\
\nuc{95}{Nb}& 34.975d & cum & & 4.47 $\pm$0.35& & 4.29 $\pm$0.38&  \\
\nuc{92}{Nb}&  10.15d & ind & 4.90$\pm$0.68 & 3.00 $\pm$0.28& & 4.17 $\pm$0.35& 3.49$\pm$0.34\\
\nuc{90}{Nb}&  14.60h & cum & 41.5$\pm$ 4.0 & 29.6 $\pm$2.1 & 24.7 $\pm$1.9 & 21.9 $\pm$2.1 & 19.5$\pm$1.8 \\
\nuc{90}{Nb}&  14.60h & m+g & 32.2$\pm$ 3.2 & 25.5 $\pm$2.0 & 20.0 $\pm$2.2 & 19.6 $\pm$1.8 &  \\
\nuc{89}{Nb}&  1.18h  & cum & 1.75$\pm$0.21 & 1.28 $\pm$0.16& & 1.24 $\pm$0.12&  \\
\nuc{89}{Nb}& 1.9h  & cum & & & & 31.2 $\pm$3.5 &  \\
\nuc{88}{Nb}& 7.8m  & ind & 7.11$\pm$0.79 & 9.25 $\pm$5.38& & 6.08 $\pm$1.55& 2.23$\pm$2.47\\
\nuc{88}{Nb}&  14.5m  & cum & 3.51$\pm$0.44 & 3.48 $\pm$0.47& 3.44 $\pm$0.39& 2.41 $\pm$0.52& 2.15$\pm$0.38\\
\nuc{89}{Zr}&  78.41h & cum & 42.8$\pm$ 4.2 & 35.9 $\pm$2.6 & 32.5 $\pm$2.4 & 31.5 $\pm$2.6 & 26.8$\pm$2.5 \\
\nuc{88}{Zr}&  83.4d  & cum & 24.0$\pm$ 2.9 & 28.4 $\pm$2.1 & 23.9 $\pm$3.2 & 23.5 $\pm$2.0 & 21.8$\pm$2.0 \\
\nuc{87}{Zr}&  1.68h  & cum & 19.8$\pm$ 2.1 & 26.7 $\pm$2.6 & 22.9 $\pm$2.0 & 18.7 $\pm$1.7 & 20.6$\pm$2.3 \\
\nuc{86}{Zr}&  16.5h  & cum & 4.93$\pm$0.49 & 9.47 $\pm$0.75& 7.85 $\pm$0.63& 7.09 $\pm$0.64& 5.12$\pm$0.56\\
\nuc{90}{Y}&  3.19h  & ind & & 0.95$\pm$0.17& 1.33 $\pm$0.17& 0.92$\pm$0.17& 1.02$\pm$0.09\\
\nuc{88}{Y}& 106.65d & cum & & 11.6 $\pm$0.8 & & 7.82 $\pm$0.75&  \\
\nuc{87}{Y}&  13.37h & cum & 25.1$\pm$ 2.4 & 25.3 $\pm$2.7 & 36.5 $\pm$2.6 & 30.8 $\pm$2.5 & 28.0$\pm$3.0 \\
\nuc{87}{Y}&  13.37h & ind & 5.38$\pm$0.92 & 11.51$\pm$1.29& 13.59$\pm$1.47& 12.10$\pm$1.21& 8.99$\pm$1.39\\
\nuc{87}{Y}&  79.8h  & cum & 24.8$\pm$ 2.6 & 39.5 $\pm$2.8 & 42.3 $\pm$3.0 & 36.4 $\pm$4.0 & 28.4$\pm$2.5 \\
\nuc{87}{Y}&  79.8h  & ind & 5.93$\pm$0.66 & 14.23$\pm$2.38& & 3.23 $\pm$2.18& 3.41$\pm$0.46\\
\nuc{86}{Y}& 48m & ind & 7.76$\pm$0.80 & 15.6 $\pm$1.2 & 13.1 $\pm$1.0 & 12.1 $\pm$1.1 & 10.4$\pm$1.0 \\
\nuc{86}{Y}&  14.74h & cum & 14.0$\pm$ 1.4 & 32.4 $\pm$2.4 & 26.8 $\pm$2.0 & 25.4 $\pm$2.1 & 21.9$\pm$2.0 \\
\nuc{86}{Y}&  14.74h & ind & 9.30$\pm$1.00 & 22.9 $\pm$1.7 & 19.5 $\pm$1.4 & 18.5 $\pm$1.6 & 17.2$\pm$1.6 \\
\nuc{85}{Y}&  4.86h  & cum & & & 4.35 $\pm$2.06& 8.21 $\pm$1.23& 8.82$\pm$2.54\\
\nuc{85}{Y}&  2.68h  & cum & 2.43$\pm$0.41 & 10.0 $\pm$1.4 & 5.70 $\pm$0.56& 4.00 $\pm$0.60& 3.34$\pm$0.59\\
\nuc{85}{Y}&  2.68h  & ind & & 4.46 $\pm$0.97& & &  \\
\nuc{84}{Y}& 40m & cum & 2.27$\pm$0.24 & 8.13 $\pm$1.04& 8.87 $\pm$0.75& 8.07 $\pm$0.78& 7.21$\pm$0.65\\
%\nuc{91}{Sr}&  9.63h  & ind & & 3.64 $\pm$0.44& & &  \\
\nuc{85}{Sr}&  67.63m & cum & 2.53$\pm$0.41 & 6.72 $\pm$0.58& 6.99 $\pm$0.75& 5.30 $\pm$0.50& 4.54$\pm$0.66\\
\nuc{85}{Sr}&  67.63m & ind & & 1.22 $\pm$0.18& 0.94$\pm$0.19& 0.86$\pm$0.13& 0.66$\pm$0.16\\
\nuc{85}{Sr}&  64.84d & cum & 12.4$\pm$ 2.6 & 41.0 $\pm$3.5 & 29.2 $\pm$4.4 & & 30.1$\pm$3.2 \\
\nuc{83}{Sr}&  32.41h & cum & & 21.6 $\pm$2.4 & 20.1 $\pm$9.4 & 19.6 $\pm$1.6 & 12.9$\pm$2.1 \\
\nuc{81}{Sr}&  22.3m  & cum & & 4.34 $\pm$0.32& 4.67 $\pm$0.46& 3.38 $\pm$0.50&  \\
\hline \end{tabular} \end{center} 
\end{table*}   \begin{table*}%[t]
Continuation of Table \ref{tc}.
\begin{center} \begin{tabular}{lccccccc}
 \hline Product& Half life & Mode& \multicolumn{4}{c}{Proton enegy (GeV)}\\
\cline{4-8} & & & 0.2 & 0.8 & 1.0 & 1.4 & 1.6\\ \hline 
\nuc{84}{Rb}&  20.26m & ind & & 3.10 $\pm$1.35& 2.88 $\pm$0.54& 3.04 $\pm$0.65& 2.99$\pm$0.93\\
\nuc{84}{Rb}&  32.77d & m+g & & 3.48 $\pm$0.28& & &  \\
\nuc{83}{Rb}&  86.2d  & cum & & 31.2 $\pm$2.5 & 28.4 $\pm$6.1 & 27.6 $\pm$2.5 & 23.6$\pm$3.3 \\
\nuc{82}{Rb}&  6.472h & cum & & 12.1 $\pm$0.9 & 11.5 $\pm$0.9 & 11.0 $\pm$0.9 & 10.3$\pm$1.0 \\
\nuc{81}{Rb}&  30.5m  & cum & 1.57$\pm$0.28 & & 20.3 $\pm$2.3 & 15.0 $\pm$2.2 & 19.6$\pm$2.7 \\
\nuc{81}{Rb}&  4.576h & cum & & 26.5 $\pm$2.6 & 23.0 $\pm$2.5 & 21.9 $\pm$2.4 & 20.5$\pm$2.4 \\
\nuc{81}{Rb}&  4.576h & ind & & & 4.30 $\pm$1.72& 8.73 $\pm$2.03& 2.29$\pm$1.95\\
\nuc{79}{Rb}&  22.9m  & cum & & 4.84 $\pm$1.34& 6.06 $\pm$0.68& 5.78 $\pm$0.58& 4.91$\pm$0.80\\
\nuc{85}{Kr}&  4.480h & cum & & 1.08 $\pm$0.12& & &  \\
\nuc{79}{Kr}&  35.04h & cum & & 14.9 $\pm$1.8 & 18.0 $\pm$3.1 & 15.8 $\pm$1.9 & 14.2$\pm$1.7 \\
\nuc{77}{Kr}&  74.4m  & cum & & 7.12 $\pm$0.55& 7.74 $\pm$0.64& 8.08 $\pm$0.65& 7.38$\pm$0.67\\
\nuc{76}{Kr}&  14.8h  & cum & & 1.88 $\pm$0.20& 2.32 $\pm$0.37& 2.04 $\pm$0.40& 1.65$\pm$0.23\\
%\nuc{82}{Br}&  35.30h & m+g & & 1.19 $\pm$0.15& 4.00 $\pm$1.51&  & 5.47$\pm$0.56\\
\nuc{77}{Br}& 57.036h & cum & & 15.4 $\pm$1.6 & 17.8 $\pm$2.5 & 19.1 $\pm$1.6 & 14.9$\pm$1.4 \\
\nuc{76}{Br}&  16.2h  & m+g & & 11.0 $\pm$1.1 & 13.6 $\pm$1.1 & 14.1 $\pm$1.4 & 13.1$\pm$1.2 \\
\nuc{76}{Br}&  16.2h  & ind & & 9.52 $\pm$1.27& 11.3 $\pm$1.0 & 12.2 $\pm$1.1 & 11.5$\pm$1.2 \\
\nuc{75}{Br}&  96.7m  & cum & & 8.18 $\pm$0.62& 9.46 $\pm$0.72& 10.26$\pm$0.87& 9.47$\pm$0.89\\
\nuc{74}{Br}& 46m & ind & & 1.60 $\pm$0.18& 0.78$\pm$0.29& 1.77 $\pm$0.29& 1.59$\pm$0.42\\
\nuc{74}{Br}&  25.4m  & cum & & & 1.83 $\pm$0.50& 18.0 $\pm$1.6 & 2.38$\pm$0.94\\
\nuc{75}{Se}& 119.770d& cum & & 13.5 $\pm$1.2 & & &  \\
\nuc{73}{Se}&  39.8m  & cum & & 3.22 $\pm$0.97& 5.01 $\pm$0.95& 5.11 $\pm$0.84& 6.38$\pm$1.06\\
\nuc{73}{Se}&  7.15h  & cum & & 5.44 $\pm$0.41& 7.19 $\pm$0.54& 8.11 $\pm$0.68& 8.07$\pm$0.75\\
\nuc{73}{Se}&  7.15h  & ind & & 3.09 $\pm$1.06& 3.65 $\pm$1.01& 3.00 $\pm$0.79& 1.69$\pm$0.91\\
\nuc{72}{Se}&  8.40d  & cum & & 1.28 $\pm$0.66& 3.26 $\pm$0.77& 3.63 $\pm$0.32&  \\
%\nuc{76}{As}& 1.0778d & ind & & 5.46 $\pm$0.57& 9.25 $\pm$1.62& 4.75 $\pm$1.55&  \\
\nuc{74}{As}&  17.77d & ind & & 2.56 $\pm$0.27& & 3.48 $\pm$0.39& 3.42$\pm$0.42\\
\nuc{72}{As}&  26.0h  & cum & & 7.90 $\pm$0.65& 11.7$\pm$1.1& 16.35$\pm$1.38& 11.9$\pm$1.15\\
\nuc{72}{As}&  26.0h  & ind & & 6.61 $\pm$0.90& 8.51 $\pm$0.70& 12.7$\pm$1.1&  \\
\nuc{71}{As}&  65.28h & cum & & 6.07 $\pm$0.48& 6.99 $\pm$0.60& 11.3$\pm$1.0& 8.90$\pm$0.98\\
\nuc{70}{As}&  52.6m  & cum & & & 5.41 $\pm$1.53& 2.25 $\pm$0.32& 2.97$\pm$0.55\\
\nuc{69}{Ge}&  39.05h & cum & & 3.46 $\pm$0.49& & 7.19 $\pm$0.75& 5.95$\pm$0.74\\
\nuc{67}{Ge}&  18.9m  & cum & & & 0.88$\pm$0.12& 1.23 $\pm$0.16& 1.17$\pm$0.17\\
%\nuc{73}{Ga}&  4.86h  & cum & & & & 1.03 $\pm$0.37&  \\
%\nuc{72}{Ga}&  14.10h & cum & & 1.09 $\pm$0.26& & & 0.37$\pm$0.28\\
\nuc{67}{Ga}& 3.2612d & cum & & 4.13 $\pm$0.70& & 9.06 $\pm$1.31& 8.72$\pm$1.07\\
\nuc{66}{Ga}&  9.49h  & cum & & 1.59 $\pm$0.36& & 4.72 $\pm$0.45& 4.87$\pm$0.74\\
\nuc{65}{Ga}&  15.2m  & cum & & & & 1.09 $\pm$0.29&  \\
\nuc{54}{Mn}& 312.12d & ind & & & & 3.71 $\pm$0.47&  \\
\nuc{52}{Mn}&  5.591d & ind & & 0.39$\pm$0.17& & 0.75$\pm$0.07&  \\
\nuc{48}{Cr}&  21.56h & ind & & 1.52 $\pm$0.13& & &  \\
\nuc{48}{V}& 15.9735d& cum & & & & 0.87$\pm$0.10&  \\
\nuc{47}{Ca}&  4.536d & cum & & 2.84 $\pm$0.70& & 2.96 $\pm$0.49&  \\
\hline  \end{tabular}
\end{center} 
\end{table*}

\begin{table*}%[t]
\caption{Experimental yields measured in \nuc{nat}{U}.}
\label{uran} 
\begin{center} \begin{tabular}{lcccccc}
 \hline Product& Half life & Mode& \multicolumn{4}{c}{Proton enegy (GeV)}\\ 
\cline{4-7} & & & 0.1 & 0.8 & 1.2 & 1.6\\ 
\hline 
\nuc{238}{Np} & 50.808h & cum & 2.46 $\pm$ 0.33 &  &  &  \\
\nuc{237}{U} & 6.75d & cum & 115 $\pm$ 12 & 114 $\pm$ 9 & 111 $\pm$ 10 & 119 $\pm$ 12 \\
\nuc{233}{Pa} & 26.967d & ind &  & 14.3 $\pm$ 1.2 & 13.5 $\pm$ 1.1 & 12.3 $\pm$ 1.1 \\
\nuc{232}{Pa} & 31.44h & cum & 3.27 $\pm$ 0.38 & 7.13 $\pm$ 1.16 & 6.46 $\pm$ 0.60 & 5.77 $\pm$ 0.58 \\
\nuc{230}{Pa} & 17.4d & ind &  & 2.38 $\pm$ 0.49 & 2.13 $\pm$ 0.38 & 1.30 $\pm$ 0.30 \\
\nuc{225}{Ac} & 10d & cum &  & 2.31 $\pm$ 0.23 & 2.30 $\pm$ 0.32 & 1.68 $\pm$ 0.21 \\
\nuc{224}{Ra} & 87.833h & cum &  & 2.40 $\pm$ 0.29 &  &  \\
\nuc{211}{Rn} & 14.6h & cum &  & 2.46 $\pm$ 0.25 & 2.09 $\pm$ 0.19 & 1.83 $\pm$ 0.37 \\
\nuc{210}{At} & 8.1h & cum &  & 3.61 $\pm$ 0.52 & 3.14 $\pm$ 0.40 & 2.60 $\pm$ 0.33 \\
\nuc{209}{At} & 5.41h & cum &  & 7.78 $\pm$ 0.85 & 7.54 $\pm$ 0.95 & 6.18 $\pm$ 0.92 \\
\nuc{208}{At} & 1.63h & cum &  & 2.72 $\pm$ 0.27 & 3.17 $\pm$ 0.73 & 3.84 $\pm$ 0.63 \\
\nuc{206}{At} & 30m & cum &  & 2.53 $\pm$ 0.33 & 3.75 $\pm$ 0.42 &  \\
\nuc{207}{Po} & 5.8h & cum &  & 7.60 $\pm$ 0.67 & 6.70 $\pm$ 0.67 & 5.03 $\pm$ 0.76 \\
\nuc{206}{Po} & 8.8d & cum &  & 8.85 $\pm$ 0.71 & 9.06 $\pm$ 0.79 & 7.30 $\pm$ 0.70 \\
\nuc{204}{Po} & 3.53h & cum &  & 4.68 $\pm$ 0.57 & 5.77 $\pm$ 0.63 & 4.39 $\pm$ 0.57 \\
\nuc{206}{Bi} & 6.243d & cum &  & 8.48 $\pm$ 0.68 &  &  \\
\nuc{205}{Bi} & 15.31d & cum &  & 6.17 $\pm$ 0.72 & 7.10 $\pm$ 0.70 & 5.87 $\pm$ 0.85 \\
\nuc{204}{Bi} & 11.22h & cum &  & 9.26 $\pm$ 0.86 & 10.3 $\pm$ 1.3 & 8.06 $\pm$ 1.09 \\
\nuc{204}{Bi} & 11.22h & ind &  & 4.59 $\pm$ 0.75 & 4.51 $\pm$ 1.16 & 3.67 $\pm$ 1.03 \\
\nuc{203}{Bi} & 11.76h & cum &  &  & 3.96 $\pm$ 0.53 & 3.82 $\pm$ 0.75 \\
\nuc{202}{Bi} & 1.72h & cum &  & 5.96 $\pm$ 0.55 & 10.1 $\pm$ 1.2 & 8.45 $\pm$ 1.55 \\
\nuc{203}{Pb} & 51.873h & cum &  & 6.19 $\pm$ 0.53 & 7.92 $\pm$ 0.66 & 6.96 $\pm$ 0.61 \\
\nuc{203}{Pb} & 51.873h & ind &  &   & 3.96 $\pm$ 0.59 & 3.14 $\pm$ 0.76 \\
\nuc{201}{Pb} & 9.33h & cum &  & 4.23 $\pm$ 0.45 & 6.68 $\pm$ 0.69 & 6.08 $\pm$ 0.67 \\
\nuc{200}{Pb} & 21.5h & cum &  & 3.39 $\pm$ 0.30 & 4.56 $\pm$ 0.76 & 5.03 $\pm$ 0.61 \\
\nuc{199}{Pb} & 90m & cum &  & 5.24 $\pm$ 1.01 &  &  \\
\nuc{200}{Tl} & 26.1h & cum &  & 4.02 $\pm$ 0.34 & 6.04 $\pm$ 0.52 & 5.41 $\pm$ 0.49 \\
\nuc{200}{Tl} & 26.1h & ind &  & 0.63 $\pm$ 0.19 & 1.51 $\pm$ 0.49 &   \\
\nuc{194}{Tl} & 32.8m & cum &  &  & 5.37 $\pm$ 0.75 & 3.39 $\pm$ 0.61 \\
\nuc{193}{Hg} & 11.8h & ind &  &  & 1.77 $\pm$ 0.38 & 1.25 $\pm$ 0.38 \\
\nuc{192}{Hg} & 4.85h & cum &  &  & 4.84 $\pm$ 0.53 & 5.85 $\pm$ 0.66 \\
\nuc{191}{Pt} & 69.6h & cum &  & 3.55 $\pm$ 0.79 & 4.24 $\pm$ 0.88 & 6.10 $\pm$ 1.13 \\
\nuc{188}{Pt} & 10.2d & cum &  &  & 4.83 $\pm$ 0.69 & 8.04 $\pm$ 0.95 \\
\nuc{184}{Ir} & 3.09h & cum &  &  &  & 6.39 $\pm$ 0.88 \\
\nuc{185}{Os} & 93.6d & cum &  & 1.91 $\pm$ 0.22 & 6.72 $\pm$ 0.62 & 9.43 $\pm$ 0.90 \\
\nuc{183}{Os} & 13h & cum &  & 3.08 $\pm$ 0.29 & 3.41 $\pm$ 0.33 & 5.55 $\pm$ 0.57 \\
\nuc{183}{Os} & 9.9h & cum &  &  & 2.82 $\pm$ 0.34 & 4.64 $\pm$ 0.46 \\
\nuc{181}{Re} & 19.9h & cum &  &  & 4.03 $\pm$ 0.70 & 10.5 $\pm$ 1.9 \\
\nuc{176}{Ta} & 8.09h & cum &  &  & 3.11 $\pm$ 0.72 & 8.79 $\pm$ 1.07 \\
\nuc{171}{Lu} & 8.24d & cum &  &  & 2.30 $\pm$ 0.26 & 6.77 $\pm$ 0.62 \\
\nuc{169}{Lu} & 34.06h & cum &  &  &  & 4.34 $\pm$ 0.41 \\
\nuc{166}{Yb} & 56.7h & cum &  &  &  & 3.51 $\pm$ 0.40 \\
\nuc{160}{Er} & 28.58h & cum &  &  &  & 4.69 $\pm$ 0.41 \\
\nuc{155}{Dy} & 9.9h & cum &  & 1.03 $\pm$ 0.13 & 1.97 $\pm$ 0.20 & 4.02 $\pm$ 0.38 \\
\nuc{151}{Pm} & 28.4h & cum & 4.57 $\pm$ 0.56 &  &  &  \\
\nuc{147}{Nd} & 10.98d & cum &  &  &  & 4.19 $\pm$ 0.58 \\
\nuc{146}{Pr} & 24.15m & cum & 16.0 $\pm$ 2.6 & 9.8 $\pm$ 1.7 & 10.7 $\pm$ 1.6 &  \\
\nuc{143}{Ce} & 33.1h & cum & 28.7 $\pm$ 3.1 & 12.3 $\pm$ 1.0 & 10.5 $\pm$ 0.9 & 9.64 $\pm$ 0.85 \\
\nuc{141}{Ce} & 32.5d & cum & 35.1 $\pm$ 3.9 & 19.1 $\pm$ 1.7 & 16.9
 $\pm$ 1.4 & 15.3 $\pm$ 1.4 \\
\hline \end{tabular} \end{center} 
\end{table*}   \begin{table*}%[t]
 Continuation of Table \ref{uran}.
 \begin{center}
\begin{tabular}{lcccccc} 
 \hline Product& Half life & Mode& \multicolumn{4}{c}{Proton enegy (GeV)}\\ 
\cline{4-7} & & & 0.1 & 0.8 & 1.2 & 1.6\\ \hline 
 \nuc{139}{Ce} & 137.64d & cum &  & 8.51 $\pm$ 0.73 & 8.30 $\pm$ 0.71 & 7.75 $\pm$ 0.72 \\
\nuc{132}{Ce} & 3.51h & cum &  &  & 1.06 $\pm$ 0.42 & 1.00 $\pm$ 0.69 \\
\nuc{142}{La} & 1.518h & cum & 26.7 $\pm$ 2.9 & 10.70 $\pm$ 0.92 & 11.8 $\pm$ 1.0 & 8.63 $\pm$ 0.83 \\
\nuc{140}{La} & 40.274h & cum & 35.4 $\pm$ 3.7 & 17.3 $\pm$ 1.4 & 14.5 $\pm$ 1.3 & 
13.1 $\pm$ 1.3 \\
\nuc{140}{La} & 40.274h & ind & 7.73 $\pm$ 0.84 & 3.22 $\pm$ 0.28 & 2.23 $\pm$ 0.18 & 1.76 $\pm$ 0.25 \\
\nuc{132}{La} & 4.8h & cum &  & 1.70 $\pm$ 0.83 & 2.72 $\pm$ 0.45 & 2.91 $\pm$ 0.60 \\
\nuc{132}{La} & 4.8h & ind &  &  & 1.66 $\pm$ 0.73 &  \\
\nuc{141}{Ba} & 18.27m & cum & 22.5 $\pm$ 4.2 &  &  &  \\
\nuc{140}{Ba} & 12.752d & cum & 27.5 $\pm$ 2.9 & 14.1 $\pm$ 1.2 & 12.3 $\pm$ 1.1 & 11.4 $\pm$ 1.0 \\
\nuc{139}{Ba} & 83.06m & cum & 43.9 $\pm$ 8.8 & 18.8 $\pm$ 3.5 & 17.5 $\pm$ 3.3 & 16.6 $\pm$ 3.3 \\
\nuc{133}{Ba} & 38.9h & ind &  &  & 3.74 $\pm$ 0.44 & 2.56 $\pm$ 0.59 \\
\nuc{131}{Ba} & 11.5d & cum &  & 4.13 $\pm$ 0.35 & 5.02 $\pm$ 0.45 & 4.95 $\pm$ 0.69 \\
\nuc{128}{Ba} & 58.32h & cum &  & 8.84 $\pm$ 1.08 & 3.26 $\pm$ 0.60 & 2.80 $\pm$ 0.55 \\
\nuc{138}{Cs} & 33.41m & cum & 27.5 $\pm$ 3.1 & 12.9 $\pm$ 1.2 & 12.0 $\pm$ 1.1 & 10.9 $\pm$ 1.3 \\
\nuc{136}{Cs} & 13.16d & m+g & 15.0 $\pm$ 1.6 & 5.55 $\pm$ 0.45 & 4.19 $\pm$ 0.54 & 3.59 $\pm$ 0.31 \\
\nuc{135}{Cs} & 53m & ind & 15.5 $\pm$ 1.7 &  &  &  \\
\nuc{132}{Cs} & 6.479d & ind &  &  & 12.4 $\pm$ 1.0 & 13.4 $\pm$ 1.2 \\
\nuc{129}{Cs} & 32.06h & cum &  & 10.39 $\pm$ 0.93 & 11.2 $\pm$ 1.6 & 11.3 $\pm$ 1.2 \\
\nuc{135}{Xe} & 9.14h & cum & 43.7 $\pm$ 4.7 & 19.7 $\pm$ 1.7 & 17.1 $\pm$ 1.4 & 16.0 $\pm$ 1.4 \\
\nuc{135}{Xe} & 9.14h & m+g & 26.0 $\pm$ 3.0 & 10.46 $\pm$ 1.13 & 4.79 $\pm$ 0.77 & 8.31 $\pm$ 0.93 \\
\nuc{135}{Xe} & 15.29m & cum & 14.6 $\pm$ 2.3 &  &  &  \\
\nuc{133}{Xe} & 52.56h & cum & 16.5 $\pm$ 1.8 &  &  &  \\
\nuc{127}{Xe} & 36.4d & cum &  & 11.5 $\pm$ 1.1 & 12.3 $\pm$ 1.0 & 11.4 $\pm$ 1.0 \\
\nuc{125}{Xe} & 16.9h & cum &  &   & 8.30 $\pm$ 0.69 & 8.16 $\pm$ 0.76 \\
\nuc{135}{I} & 6.57h & cum & 17.6 $\pm$ 2.1 & 9.20 $\pm$ 0.79 & 8.01 $\pm$ 0.68 & 7.66 $\pm$ 0.73 \\
\nuc{134}{I} & 52.5m & cum & 34.0 $\pm$ 3.9 &  &  &  \\
\nuc{133}{I} & 20.8h & cum & 37.6 $\pm$ 4.2 & 17.1 $\pm$ 1.5 & 15.4 $\pm$ 1.3 & 14.7 $\pm$ 1.4 \\
\nuc{131}{I} & 8.021d & cum & 51.5 $\pm$ 5.7 & 21.0 $\pm$ 1.7 & 18.2 $\pm$ 1.5 & 16.4 $\pm$ 1.4 \\
\nuc{130}{I} & 12.36h & cum & 17.0 $\pm$ 1.8 & 8.44 $\pm$ 0.69 & 6.41 $\pm$ 0.51 & 4.99 $\pm$ 0.50 \\
\nuc{126}{I} & 13.11d & ind &  & 9.66 $\pm$ 1.19 & 7.56 $\pm$ 0.98 & 6.63 $\pm$ 0.91 \\
\nuc{124}{I} & 4.18d & ind &  & 7.93 $\pm$ 0.85 & 8.18 $\pm$ 0.66 & 6.07 $\pm$ 0.55 \\
\nuc{132}{Te} & 76.896h & cum & 20.9 $\pm$ 2.6 & 11.6 $\pm$ 1.1 & 11.0 $\pm$ 1.0 & 10.71 $\pm$ 1.07 \\
\nuc{131}{Te} & 25m & cum & 26.7 $\pm$ 3.1 & 9.92 $\pm$ 1.45 & 9.41 $\pm$ 1.29 & 10.07 $\pm$ 1.13 \\
\nuc{131}{Te} & 30h & cum & 11.1 $\pm$ 1.4 & 4.27 $\pm$ 0.55 & 3.22 $\pm$ 0.55 & 2.61 $\pm$ 0.37 \\
\nuc{119}{Te} & 4.7d & ind &  & 2.41 $\pm$ 0.20 & 3.61 $\pm$ 0.29 & 3.31 $\pm$ 0.33 \\
\nuc{119}{Te} & 16.03h & cum &  & 0.59 $\pm$ 0.10 & 1.89 $\pm$ 0.16 & 2.38 $\pm$ 0.23 \\
\nuc{130}{Sb} & 39.5m & cum & 5.66 $\pm$ 0.72 &  &  &  \\
\nuc{129}{Sb} & 4.4h & cum & 6.54 $\pm$ 0.90 &  &  &  \\
\nuc{128}{Sb} & 9.01h & cum & 11.1 $\pm$ 1.2 & 4.95 $\pm$ 0.63 & 4.02 $\pm$ 0.74 & 3.28 $\pm$ 0.38 \\
\nuc{127}{Sb} & 3.85d & cum & 29.8 $\pm$ 3.3 & 11.7 $\pm$ 1.0 & 9.48 $\pm$ 0.84 & 8.37 $\pm$ 0.79 \\
\nuc{126}{Sb} & 12.46d & cum & 17.8 $\pm$ 1.9 & 7.68 $\pm$ 0.61 & 5.99 $\pm$ 0.47 & 4.98 $\pm$ 0.42 \\
\nuc{124}{Sb} & 60.2d & cum & 16.6 $\pm$ 1.8 & 12.6 $\pm$ 1.1 & 9.57 $\pm$ 1.12 & 7.52 $\pm$ 0.65 \\
\nuc{122}{Sb} & 65.371h & m+g & 6.15 $\pm$ 0.66 & 14.1 $\pm$ 1.2 & 11.3 $\pm$ 
0.9 & 9.29 $\pm$ 0.81 
\\
\nuc{120}{Sb} & 5.76d & ind &  & 9.16 $\pm$ 0.73 & 8.33 $\pm$ 0.65 & 6.72 $\pm$ 0.57 \\
\nuc{118}{Sb} & 5h & ind &  & 5.09 $\pm$ 0.45 & 5.96 $\pm$ 0.52 & 5.40 $\pm$ 0.50 \\
\nuc{128}{Sn} & 59.1m & cum & 5.75 $\pm$ 1.06 &  &  &  \\
\nuc{125}{Sn} & 9.64d & cum & 9.60 $\pm$ 3.14 &  &  &  \\
\nuc{123}{Sn} & 40.06m & cum & 20.7 $\pm$ 2.3 & 5.35 $\pm$ 0.57 & 4.91 $\pm$ 0.60 & 4.82 $\pm$ 0.57 \\
\nuc{117}{In} & 43.2m & cum &  &  & 30.7 $\pm$ 4.0 & 26.7 $\pm$ 3.6 \\
\nuc{114}{In} & 49.51d & cum &  & 10.6 $\pm$ 1.0 & 10.8 $\pm$ 1.0 & 8.15 $\pm$ 1.03 \\
\nuc{111}{In} & 67.318h & cum &  & 3.36 $\pm$ 0.34 & 5.91 $\pm$ 0.70 & 6.67 $\pm$ 0.62 \\
\hline  \end{tabular}\end{center} 
 \end{table*}  
 \begin{table*}%[t]
 Continuation of Table \ref{uran}. 
\begin{center}\begin{tabular}{lcccccc} 
 \hline Product& Half life & Mode& \multicolumn{4}{c}{Proton enegy (GeV)}\\ 
\cline{4-7} & & & 0.1 & 0.8 & 1.2 & 1.6\\ 
\hline \nuc{117}{Cd} & 2.490h & cum & 15.9 $\pm$ 2.2 & 5.34 $\pm$ 2.13 & 5.37 $\pm$ 0.59 & 4.24 $\pm$ 0.65 \\
\nuc{117}{Cd} & 3.360h & cum & 18.1 $\pm$ 2.3 & 12.6 $\pm$ 1.1 & 10.0 $\pm$ 0.9 & 8.09 $\pm$ 1.05 \\
\nuc{115}{Cd} & 53.460h & cum & 54.3 $\pm$ 7.8 & 34.7 $\pm$ 2.7 & 27.4
 $\pm$ 2.2 & 21.8 $\pm$ 2.2 
\\
\nuc{115}{Cd} & 53.46h & ind & 29.9 $\pm$ 10.3 &  &  &  \\
\nuc{111}{Cd} & 48.6m & ind &  &  & 9.75 $\pm$ 1.12 & 8.49 $\pm$ 1.03 \\
\nuc{115}{Ag} & 20m & cum & 24.5 $\pm$ 8.6 &  &  &  \\
\nuc{113}{Ag} & 5.37h & cum & 70.6 $\pm$ 7.7 & 53.5 $\pm$ 4.7 & 49.6 $\pm$ 4.2 & 38.4 $\pm$ 4.1 \\
\nuc{112}{Ag} & 3.13h & cum & 70.7 $\pm$ 11.2 & 70.2 $\pm$ 10.0 & 56.5 $\pm$ 8.0 & 46.9 $\pm$ 6.8 \\
\nuc{112}{Ag} & 3.13h & ind & 7.56 $\pm$ 1.52 & 26.4 $\pm$ 3.8 & 20.1 $\pm$ 2.9 & 14.8 $\pm$ 2.3 \\
\nuc{111}{Ag} & 7.45d & cum & 62.5 $\pm$ 7.8 &  &  &  \\
\nuc{111}{Ag} & 7.45d & ind & 51.3 $\pm$ 7.1 &  &  &  \\
\nuc{110}{Ag} & 249.79d & ind &  & 12.2 $\pm$ 1.0 & 11.8 $\pm$ 1.0 & 9.21 $\pm$ 0.91 \\
\nuc{106}{Ag} & 8.28d & ind &  & 1.68 $\pm$ 0.18 & 2.93 $\pm$ 0.31 & 3.14 $\pm$ 0.42 \\
\nuc{105}{Ag} & 41.29d & cum &  &  & 1.63 $\pm$ 0.43 & 4.53 $\pm$ 0.63 \\
\nuc{112}{Pd } & 21.03h & cum & 58.5 $\pm$ 7.6 & 43.9 $\pm$ 6.2 & 36.5 $\pm$ 5.2 & 32.1 $\pm$ 4.7 \\
\nuc{111}{Pd } & 5.5h & cum & 11.2 $\pm$ 1.8 & 11.4 $\pm$ 1.7 &  &  \\
\nuc{107}{Rh} & 21.7m & cum & 77.3 $\pm$ 11.0 &  &  &  \\
\nuc{106}{Rh} & 2.183h & ind &  & 13.1 $\pm$ 1.4 & 14.0 $\pm$ 2.0 & 11.8 $\pm$ 1.1 \\
\nuc{105}{Rh} & 35.36h & cum & 76.1 $\pm$ 8.4 & 75.2 $\pm$ 6.7 & 68.1 $\pm$ 7.2 & 55.3 $\pm$ 7.8 \\
\nuc{105}{Rh} & 35.36h & ind & 15.3 $\pm$ 3.4 &  &  &  \\
\nuc{101}{Rh} & 4.34d & ind &  & 1.72 $\pm$ 0.22 & 3.58 $\pm$ 0.42 & 4.26 $\pm$ 0.52 \\
\nuc{100}{Rh} & 20.8h & cum &  & 0.26 $\pm$ 0.15 & 1.86 $\pm$ 0.44 & 2.48 $\pm$ 0.64 \\
\nuc{105}{Ru} & 4.44h & cum & 60.8 $\pm$ 6.4 & 51.4 $\pm$ 5.1 & 42.8 $\pm$ 3.8 & 35.6 $\pm$ 3.4 \\
\nuc{103}{Ru} & 39.26d & cum & 70.7 $\pm$ 7.6 & 73.0 $\pm$ 6.3 & 65.6 $\pm$ 5.1 & 55.5 $\pm$ 5.0 \\
\nuc{104}{Tc} & 18.4m & cum & 63.1 $\pm$ 8.8 & 39.9 $\pm$ 3.8 & 35.1 $\pm$ 3.1 & 35.3 $\pm$ 5.3 \\
\nuc{101}{Tc} & 14.2m & cum & 78.3 $\pm$ 12.7 & 80.5 $\pm$ 18.4 & 88.7 $\pm$ 17.1 &  \\
\nuc{101}{Tc} & 14.2m & ind &  & 43.3 $\pm$ 22.7 & 59.6 $\pm$ 19.9 &  \\
\nuc{99}{Tc} & 6.01h & cum & 66.6 $\pm$ 7.3 & 62.8 $\pm$ 5.4 & 57.3 $\pm$ 4.8 & 50.7 $\pm$ 4.5 \\
\nuc{99}{Tc} & 6.01h & ind & 0.00 $\pm$ 0.12 & 1.72 $\pm$ 0.22 & 2.90 $\pm$ 0.30 & 3.82 $\pm$ 0.45 \\
\nuc{96}{Tc} & 4.28d & m+g & 0.69 $\pm$ 0.19 &  & 3.96 $\pm$ 0.54 & 5.12 $\pm$ 1.01 \\
\nuc{95}{Tc} & 20.0h & ind &  & 0.67 $\pm$ 0.11 & 1.89 $\pm$ 0.16 & 2.91 $\pm$ 0.27 \\
\nuc{101}{Mo} & 14.61m & cum & 66.1 $\pm$ 8.0 & 37.1 $\pm$ 7.4 & 29.1 $\pm$ 5.5 &   \\
\nuc{99}{Mo} & 65.94h & cum & 66.3 $\pm$ 7.2 & 70.0 $\pm$ 5.7 & 58.7 $\pm$ 4.9 & 49.3 $\pm$ 5.1 \\
\nuc{93}{Mo} & 6.85h & ind &  &  & 1.17 $\pm$ 0.12 & 1.68 $\pm$ 0.16 \\
\nuc{98}{Nb} & 51.3m & ind & 9.15 $\pm$ 1.17 & 13.8 $\pm$ 1.3 & 8.9
$\pm$ 0.8 & 11.4 $\pm$ 1.3 \\
\nuc{96}{Nb} & 23.35h & ind & 1.84 $\pm$ 0.21 & 16.9 $\pm$ 1.4 & 17.9
$\pm$ 1.6 & 16.1 $\pm$ 1.4 \\
\nuc{95}{Nb} & 34.975d & cum &  & 74.2 $\pm$ 6.1 & 67.5 $\pm$ 5.3 &
61.5 $\pm$ 5.3 \\
\nuc{95}{Nb} & 34.975d & ind &  & 23.7 $\pm$ 3.3 & 66.2 $\pm$ 5.3 &
17.0 $\pm$ 1.5 \\
\nuc{90}{Nb} & 14.6h & cum &  &  & 1.22 $\pm$ 0.12 & 1.63 $\pm$ 0.33 \\
\nuc{97}{Zr} & 16.91h & cum & 57.9 $\pm$ 6.1 & 35.4 $\pm$ 2.8 & 31.2 $\pm$ 2.5 & 28.1 $\pm$ 3.1 \\
\nuc{95}{Zr} & 64.02d & cum & 59.8 $\pm$ 6.8 & 50.5 $\pm$ 4.2 & 43.9 $\pm$ 3.5 & 40.0 $\pm$ 3.9 \\
\nuc{89}{Zr} & 78.41h & cum &   & 2.71 $\pm$ 0.22 & 5.33 $\pm$ 0.42 & 6.61 $\pm$ 0.57 \\
\nuc{94}{Y} & 18.7m & ind & 47.8 $\pm$ 6.1 & 27.0 $\pm$ 3.2 & 22.8 $\pm$ 3.2 & 22.4 $\pm$ 2.8 \\
\nuc{93}{Y} & 10.18h & cum & 42.5 $\pm$ 7.9 & 31.7 $\pm$ 6.4 & 38.1 $\pm$ 7.9 & 20.1 $\pm$ 4.6 \\
\nuc{92}{Y} & 3.54h & cum & 48.1 $\pm$ 7.8 & 42.2 $\pm$ 6.0 & 35.5 $\pm$ 5.5 & 29.0 $\pm$ 4.9 \\
\nuc{92}{Y} & 3.54h & ind & 13.1 $\pm$ 4.4 & 10.9 $\pm$ 2.3 &  &  \\
\nuc{91}{Y} & 49.71m & cum & 25.2 $\pm$ 3.1 & 34.4 $\pm$ 2.9 & 33.9 $\pm$ 2.9 & 30.9 $\pm$ 3.1 \\
\nuc{91}{Y} & 49.71m & ind &  & 13.1 $\pm$ 1.3 & 15.9 $\pm$ 1.6 & 14.9 $\pm$ 2.0 \\
\nuc{90}{Y} & 3.19h & ind &  & 12.0 $\pm$ 1.1 & 14.1 $\pm$ 1.5 & 12.5 $\pm$ 1.1 \\
\nuc{88}{Y} & 106.65d & cum &  & 6.38 $\pm$ 0.68 & 10.9 $\pm$ 0.9 &
11.4 $\pm$ 1.0 \\
\nuc{87}{Y} & 13.37h & cum &  & 3.60 $\pm$ 0.32 & 5.65 $\pm$ 0.46 & 8.88 $\pm$ 0.93 \\
\nuc{87}{Y} & 79.8h & cum &  & 4.35 $\pm$ 0.37 & 8.29 $\pm$ 0.68 & 9.63 $\pm$ 0.84 \\
\nuc{86}{Y} & 14.74h & cum &  & 0.73 $\pm$ 0.09 & 2.07 $\pm$ 0.23 & 2.65 $\pm$ 0.45 \\
\hline  \end{tabular} \end{center} 
 \end{table*}   

\begin{table*} [t] 
Continuation of Table \ref{uran}. 
\begin{center}\begin{tabular}{lcccccc}
 \hline Product& Half life & Mode& \multicolumn{4}{c}{Proton enegy (GeV)}\\ 
\cline{4-7} & & & 0.1 & 0.8 & 1.2 & 1.6\\ \hline 
 \nuc{92}{Sr} & 2.71h & cum & 37.3 $\pm$ 4.5 & 27.8 $\pm$ 2.8 & 23.3 $\pm$ 1.8 & 19.8 $\pm$ 2.1 \\
\nuc{91}{Sr} & 9.63h & cum & 46.0 $\pm$ 4.9 & 33.9 $\pm$ 3.2 & 28.5 $\pm$ 3.7 & 24.4 $\pm$ 3.4 \\
\nuc{89}{Rb} & 15.15m & cum & 34.9 $\pm$ 5.2 & 20.5 $\pm$ 2.8 & 19.8 $\pm$ 2.9 & 18.0 $\pm$ 3.2 \\
\nuc{86}{Rb} & 18.631d & m+g &  & 12.2 $\pm$ 1.1 & 15.3 $\pm$ 1.3 & 13.2 $\pm$ 1.3 \\
\nuc{83}{Rb} & 86.2d & cum &  & 3.54 $\pm$ 0.37 & 7.45 $\pm$ 0.69 & 9.28 $\pm$ 0.91 \\
\nuc{88}{Kr} & 2.84h & cum & 19.9 $\pm$ 2.2 & 11.5 $\pm$ 1.1 & 10.1
$\pm$ 1.0 & 8.04 $\pm$ 0.78 \\
\nuc{87}{Kr} & 76.3m & cum & 27.2 $\pm$ 3.7 & 18.3 $\pm$ 2.1 & 14.9 $\pm$ 1.7 & 12.0 $\pm$ 1.5 \\
\nuc{85}{Kr} & 4.48h & cum & 18.3 $\pm$ 2.3 & 13.0 $\pm$ 1.8 & 11.9 $\pm$ 1.1 & 9.9 $\pm$ 1.3 \\
\nuc{84}{Br} & 31.8m & cum & 12.7 $\pm$ 3.0 &  &  &  \\
\nuc{82}{Br} & 35.3h & m+g &  & 10.7 $\pm$ 0.9 & 11.4 $\pm$ 0.9 &
10.3 $\pm$ 0.9 \\
\nuc{77}{Br} & 57.036h & cum &  & 0.49 $\pm$ 0.18 & 0.76 $\pm$ 0.17 & 1.42 $\pm$ 0.38 \\
\nuc{76}{Br} & 16.2h & cum &  & 0.33 $\pm$ 0.95 & 0.82 $\pm$ 0.43 &  \\
\nuc{83}{Se} & 22.3m & cum & 6.57 $\pm$ 1.15 &  &  &  \\
\nuc{75}{Se} & 119.77d & cum &  &  & 2.34 $\pm$ 0.20 & 2.92 $\pm$ 0.32 \\
\nuc{78}{As} & 1.512h & cum &  & 10.23 $\pm$ 1.45 & 9.00 $\pm$ 1.28 & 6.25 $\pm$ 0.95 \\
\nuc{76}{As} & 25.867h & ind &  & 4.92 $\pm$ 1.03 & 6.27 $\pm$ 0.79 & 7.52 $\pm$ 0.86 \\
\nuc{74}{As} & 17.77d & ind &  & 2.11 $\pm$ 0.24 & 4.53 $\pm$ 0.49 & 5.33 $\pm$ 0.63 \\
\nuc{78}{Ge} & 88m & cum & 1.96 $\pm$ 0.60 &  &  &  \\
\nuc{77}{Ge} & 11.3h & cum & 1.14 $\pm$ 0.16 & 4.17 $\pm$ 0.40 &  &  \\
\nuc{73}{Ga} & 4.86h & cum & 1.28 $\pm$ 0.19 &  &  &  \\
\nuc{72}{Zn} & 46.5h & cum & 1.10 $\pm$ 0.29 & 2.74 $\pm$ 0.25 & 2.71 $\pm$ 0.25 & 2.38 $\pm$ 0.23 \\
\nuc{71}{Zn} & 3.96h & cum &  & 2.59 $\pm$ 0.31 & 3.51 $\pm$ 0.34 & 2.80 $\pm$ 0.32 \\
\nuc{69}{Zn} & 13.76h & ind &  & 2.72 $\pm$ 0.24 & 4.30 $\pm$ 0.36 & 3.92 $\pm$ 0.37 \\
\nuc{58}{Co} & 70.916d & m+g &  &  & 0.45 $\pm$ 0.06 & 0.94 $\pm$ 0.09 \\
\nuc{59}{Fe} & 44.503d & cum &  & 2.85 $\pm$ 0.85 & 3.21 $\pm$ 0.37 & 3.60 $\pm$ 0.48 \\
\nuc{48}{Sc } & 43.67h & cum &  & 0.45 $\pm$ 0.05 & 1.05 $\pm$ 0.17 & 1.56 $\pm$ 0.22 \\
\nuc{46}{Sc } & 83.81d & m+g &  &  &  & 1.06 $\pm$ 0.11 \\
\nuc{28}{Mg} & 20.91h & cum &  &  & 0.33 $\pm$ 0.05 & 0.67 $\pm$ 0.07 \\
\nuc{24}{Na} & 14.959h & cum &  &  & 0.97 $\pm$ 0.09 & 2.06 $\pm$ 0.18 \\
\hline  \end{tabular}
\end{center} 
\end{table*}

with its standard deviation $S(\langle H \rangle)$
\begin{equation}
S(\langle H \rangle) = 10^{\sqrt{a}} ,\end{equation}
where $$a=\langle \left( \Bigl| ~lg\left( \frac{\sigma_{cal,i}}{\sigma_{exp,i}}
\right)\Bigl|\;-\;lg(\langle H \rangle) \right)^2\rangle ,$$
{\noindent
where $\langle~\rangle$ designates averaging over all of the comparison
events ($i=1...N_S$, where $N_S$ is the
number of experimental and simulated events used for comparison).}

The mean squared deviation factor $\langle H \rangle$
with its standard deviation $S(\langle H \rangle)$
define the interval  $[\langle H \rangle / S(\langle H \rangle), \langle H
\rangle \times S(\langle H \rangle) ]$
which covers about two thirds of all the simulation-to-experiment ratios.

Table \ref{comp_t} presents information concerning
the predictive power of each of the
codes for  high and low energies, namely, the total
number of measured yields $N_E$,
the number of the latter that was chosen to be used in the comparison
with simulated data $N_G$,
the total number of simulated products that
can be compared with the
data $N_S$,
the two numbers of ``coincidences" between simulated and experimental values
$N_{C_{1.3}}, N_{C_{2.0}}$, %$N_C$,
and the mean deviation $\langle H \rangle $ with its
$S(\langle H \rangle )$ of simulation results from experimental data.

\begin{table*} [t]
\caption{Statistics of simulation-to-experiment comparisons of the yields of
the measured reaction products for \nuc{99}{Tc} and \nuc{nat}{U}.}
\label{comp_t}
\begin{center}
\begin{tabular}{lccccccc}
\hline
& & & & & & & \\
& & & &{\large \nuc{99}{Tc}}&&&\\
& \multicolumn{3}{c}{E$_p$ = 0.2 GeV}
&& \multicolumn{3}{c}{E$_p$ = 0.8, 1.0, 1.4, and 1.6 GeV}\\
& \multicolumn{3}{c}{N$_E$ = 39, N$_G$ = 25}
&& \multicolumn{3}{c}{N$_E$ = 284, N$_G$ = 200}\\
\quad Code &N$_{1.3}$/N$_{2.0}$/N$_S$&$\langle H \rangle$&$S(\langle H \rangle)$
&&N$_{1.3}$/N$_{2.0}$/N$_S$&$\langle H \rangle$&$S(\langle H \rangle)$\\
& & & & & & & \\
\hline 
CEM95&8/14/25&3.84&2.86&&82/143/195&2.08&1.74\\
LAHET(BERTINI)&8/18/25&1.84&1.48&&77/155/196&1.80&1.51\\
LAHET(ISABEL)&9/20/25&1.83&1.50&&53/83/97$^1$&1.75&1.57\\
INUCL&6/9/25&2.83&1.86&&42/101/194&2.73&1.93\\
HETC&5/12/22&3.62&2.58&&52/99/188&4.04&2.96\\
CASCADE&7/15/25&2.82&2.00&&58/111/195&2.59&1.92\\
ALICE(Fermi)&6/18/25&2.16&1.69&&--&--&--\\
ALICE(Kataria)&8/11/25&2.43&1.75&&--&--&--\\
YIELDX&9/19/25&2.13&1.71&&75/154/197&2.13&1.78\\
Foshina et al.&3/4/25&5.48&2.16&&34/76/198&2.99&1.84\\
%GNASH&2/15/14&3.38&2.21&&--&--&--\\
\hline 
& & & & & & & \\
&&&&{\large \nuc{nat}{U}}&&&\\
& \multicolumn{3}{c}{E$_p$ = 0.1 GeV}
&& \multicolumn{3}{c}{E$_p$ = 0.8, 1.2, and 1.6 GeV}\\
& \multicolumn{3}{c}{N$_E$ = 83, N$_G$ = 63}
&& \multicolumn{3}{c}{N$_E$ = 418, N$_G$ = 328}\\
\quad Code&N$_{1.3}$/N$_{2.0}$/N$_S$&$\langle H \rangle$&$S(\langle H \rangle)$
&&N$_{1.3}$/N$_{2.0}$/N$_S$&$\langle H \rangle$&$S(\langle H \rangle)$\\
& & & & & & & \\
\hline 
CEM95& 0/2/2&1.62&1.02&&16/42/78&2.89&2.11\\
LAHET(BERTINI)& 15/47/63&2.29&1.82&&76/186/326&2.25&1.62\\
LAHET(ISABEL)&19/45/63&2.32&1.88&&26/66/98$^2$&1.99&1.57\\
INUCL&10/16/61&7.25&3.68&&49/141/296&4.48&3.00\\
HETC&2/2/3&2.45&1.99&&9/15/61&10.5&4.3\\
CASCADE&0/1/3&18.1&6.0&&10/16/25$^2$&3.13&2.45\\
\hline
& & & & & & & \\
\multicolumn{7}{l}{$^1$ - only 0.8 and 1.0 GeV simulated.}\\
\multicolumn{7}{l}{$^2$ - only 0.8 GeV simulated.}
\end{tabular} 
\end{center} 
\end{table*}

\vspace{0.3cm}
\noindent IV. CONCLUSION
\vspace{0.1cm}

This study is the first step in our work
on the fissible targets that are
of interest for accelerator-driven facilities.
%Our 2-year long activities are summarized in Table \ref{sum}.
In total, about 820 reaction data have been measured.
Table \ref{comp_t} reflects the predictive
power of the simulation codes used here.

It should be noted that 
the CEM95, HETC, and CASCADE codes
do not calculate the process of fission itself, and do
not provide fission fragments and a further possible evaporation of
particles from them. When, during the Monte Carlo simulation of a
compound stage of a reaction using evaporation and fission widths, 
these codes have to simulate a fission, 
they simply remember this event (that 
permits them to calculate fission cross section and fissility)
and finish the calculation of this event 
without a real subsequent calculation of fission fragments and a further 
possible evaporation of particles from them.
Therefore, the results from 
CEM95, HETC, and CASCADE shown in Tab. 4
reflect the contribution to
the total yields of the nuclides only from deep spallation processes of 
successive emission of particles from the target, 
but do not contain fission products. 
To be able to describe 
nuclide production in the fission region, these codes have to be extended
by incorporating a model of high energy fission. 

At present, analyzing our theoretical results
we  come to the conclusion that the differences among some
theoretical
yields predicted by different codes can sometimes be very significant.
This is a strong indication
that all of the codes have to be further improved before they can become
reliable predictive tools.
Therefore, the relevant experiments have to be extended.

We continue our measurements and expect, for instance, to have in the 
following year new data for 
\nuc{nat}{U} thin target irradiated by 0.2 GeV protons, 
\nuc{99}{Tc} irradiated by 0.1 and 1.2 GeV protons, 
and \nuc{232}{Th} irradiated by 0.1, 0.2, 0.8, 1.2, and 1.6 GeV protons.

The Cascade-Exciton Model (CEM) of nuclear reactions
is under development at present 
at LANL and a recent
version of the code named CEM97 with a much better predictive power than of
the code CEM95 has been already realized$^{16}$.% \ref{cem97}. 
We began to develop a fission 
model appropriate for the CEM to describe mass,
charge, energy and angular distributions of fission fragments with a 
possible further evaporation of particle from fragments (or even
sequental fission)$^{17}$ % \ref{cemfis} 
and we plan to incorporate this model into
a newer version 
of the CEM code, CEM99. The U-data of the present work and 
from future planed measurements will be used 
together with other available data to fix the parameters of CEM99.

\vspace{0.5cm}
\noindent ACKNOWLEDGMENTS
\vspace{0.1cm}

The work was performed under 
%initiated under the ISTC Project \# 017 and was
%completed under %the Feasibility Study Stage of
the ISTC Project \# 839 %.
%This study 
and
was partially supported by the U. S. Department of Energy.

We thank C. H. Tsao for providing us with his 
latest version of the YIELDX code and
F. E. Chukreev, V. A. Vukolov, N. V. Stepanov, 
and M. B. Chadwick for useful co-operation in
the work.

\vspace{0.4cm}
\noindent REFERENCES
\vspace{0.2cm}

\begin{enumerate}

\vspace*{-0.2cm}
\item
G. S. Bauer, ``Research and Development for Molten Heavy
Metal Targets",
{\it Proc. 2nd Int. Conf. on Accelerator-Driven
Transmutation Technologies and Applications}
Kalmar, Sweden, June 3-7, 1996,
ed. H. Cond\'e (Uppsala University Press, 1997), Vol. 2, pp. 803-814.

\vspace*{-0.2cm}
\item
A. J. Koning,
``Review of High Energy Data and Model Codes for
Accelerator-Based Transmutation,"
ECN-C-93-005, Petten (1993).

\vspace*{-0.2cm}
\item
Yu. E. Titarenko, O. V. Shvedov, V. F. Batyaev, 
E. I. Karpikhin, V. M. Zhivun, R. D. Mulambetov, A. N. Sosnin,
S. G. Mashnik, R. E. Prael, M. B. Chadwick,
T. A. Gabriel, and M. Blann
``Experimental and Theoretical Study of the Yields of Radioactive Product  
Nuclei  in 99-Tc  Thin  Targets Irradiated by  100-2600 MeV  Protons",
{\em Proc. 3rd Int. Conf. on Accelerator Driven Transmutation Technologies
and Applications}, Praha (Pruhonice), June 7-11, 1999, Czech Republic.

\vspace*{-0.2cm}
\item
Yu. E. Titarenko, O. V. Shvedov, V. F. Batyaev, 
E. I. Karpikhin,  V. M. Zhivun,  A. B. Koldobsky, 
R. D. Mulambetov, A. N. Sosnin,
Yu. N. Shubin, A. V. Ignatyuk, V. P. Lunev,
S. G. Mashnik, R. E. Prael, T. A. Gabriel, and M. Blann
``Experimental and Theoretical  Study of the  Yields of Radionuclides 
Produced in nat-U Thin Targets Irradiated by 100 and 800 MeV Protons",
ibid. 
%3$^{rd}$ Int. Conf. on Accelerator Driven Transmutation Technologies
%and Applications, June 7-11, 1999, Praha, Czech Republic.

\vspace*{-0.2cm}    % 5
\item
Yu. E. Titarenko, O. V. Shvedov, M. M. Igumnov, 
S. G. Mashnik, 
E. I. Karpikhin, V. D. Kazaritsky, V. F. Batyaev,
A. B. Koldobsky, V. M. Zhivun, A. N. Sosnin, R. E. Prael,
M. B. Chadwick, T. A. Gabriel, and M. Blann,
Experimental and Computer Simulation Study of the Radionuclides Produced in
Thin $^{209}$Bi Targets by 130 MeV and 1.5 GeV Proton-Induced Reactions.
{\em Nucl. Instr. Meth.}, {\bf A414} (1998) 73.

\vspace*{-0.2cm}        %6
\item
K. K. Gudima, S. G. Mashnik, and V. D. Toneev,
{\it Nucl. Phys.}, {\bf A401}, (1983) 329;
S.G. Mashnik,
{\it Bull. Russian Ac. Sci.: Physics}, {\bf 60}, (1996) 58;
S. G. Mashnik, A. J. Sierk, O. Bersillon, and T. Gabriel,
LANL Reports LA-UR-97-2905, Los Alamos (1997) and
{\em Nucl. Instr. Meth.}, {\bf A414} (1998) 68.

\vspace*{-0.2cm}
\item
V. S. Barashenkov et al.,
{\it Sov. J. Nucl. Phys.}, {\bf 39} (1984) 715; %];
{\it Nucl. Phys.}, {\bf A338} (1980) 413;
JINR Report R2-85-173, Dubna (1985).

\vspace*{-0.2cm}
\item
N. V. Stepanov,
ITEP Preprints ITEP-91 (1983), ITEP-129 (1985), ITEP-81 (1987),
ITEP-55-88, Moscow (1988) [in Russian].

\vspace*{-0.2cm}
\item
T. W. Armstrong and K. C. Chandler,
{\it Nucl. Sci. Eng.}, {\bf 49} (1972) 110. % and references therein.

\vspace*{-0.2cm}
\item
R. E. Prael and H. Lichtenstein,
Los Alamos National Laboratory Report LA-UR-89-3014 (1989).
%
%\vspace*{-0.2cm}
%\item
%M. B. Chadwick and P. G. Young,
%{\it Phys. Rev.}, {\bf C47} (1993) 2255.

\vspace*{-0.2cm}
\item
M. Blann,
 ``New Precompound Decay Model", {\it Phys. Rev.}, {\bf C54} (1996) 1341.

\vspace*{-0.2cm}
\item
C. H. Tsao, priv. comm.; R. Silberberg and C. H. Tsao,
{\it Astrophys. J.}, {\bf 220} (1973) 315; ibid p. 335.

\vspace*{-0.2cm}
\item
M. Foshina et al.,
{\it Radiochimica Acta}, {\bf 35} (1984) 121.

\vspace*{-0.2cm}
\item
R. Michel and P. Nagel,
{\em International Codes and Model Intercomparison for Intermediate
Energy Activation Yields},
NEA/OECD, NSC/DOC(97)-1, Paris (1997).

\vspace*{-0.2cm}
\item
Yu.E. Titarenko et al.,
``Experimental and Computer Simulation Study of Radionuclide Formation 
in the ADT Materials Irradiated with Intermediate Energy Protons",
{\em Proc. 2nd Int. Topical Meeting on Nuclear Applications of Accelerator 
Technology},
September 20-23, 1998, Gatlinburg, Tennessee.
%, 164--171.

\vspace*{-0.2cm}
\item
S. G. Mashnik and A. J. Sierk,
``Improved Cascade-Exciton Model of Nuclear Reactions,"
{\em Proc. Fourth Workshop on Simulating Accelerator Radiation Envirinments 
(SARE4)}, Knoxville, TN, USA, September 14-16, 1998, pp. 29-51 and
LANL Report LA-UR-98-5999, Los Alamos (1998);
H. G. Hughes, K. J. Adams, M. B. Chadwick, J. C. Comly, L. J. Cox,
H. W. Egdorf, S. C. Frankle, J. S. Hendricks, R. C. Little, R. MacFarlane,
S. G. Mashnik, R. E. Prael, A. J. Sierk, L. S. Waters, M. C. White,
P. G. Young, F. Gallmeier, and E. C. Snow,
``MCNPX for Neutron-Proton Transport,"
LANL Report LA-UR-99-1935, Los Alamos (1999), submited to 
{\em M\&C'99}, September 27-30, 1999, Madrid, Spain.

\vspace*{-0.2cm}
\item
A. J. Sierk and S. G. Mashnik,
``Modeling Fission in the Cascade-Exciton Model,"
{\em Proc. Fourth Workshop on Simulating Accelerator Radiation Environments 
(SARE4)}, Knoxville, TN, USA, September 14-16, 1998, pp. 53-67 and
LANL Report LA-UR-98-5998, Los Alamos (1998).

\end{enumerate}
\end{document}